\documentclass[a4paper]{jpconf}

\usepackage{graphicx}
\begin{document}

\title{Probing the magnetic phases in the Ni-V alloy close to the disordered ferromagnetic quantum critical point with $\mu$SR
}

\author{A Schroeder$^1$, R Wang$^1$, P J Baker$^2$, F L Pratt$^2$, S J Blundell$^3$, T Lancaster$^{3,}$\footnote[4]{Present address: Durham University, Centre for Materials Physics, South Road, Durham DH1 3LE, UK}, I Franke$^3$, J S M{\"o}ller$^3$ }

\address{$^1$ Department of Physics, Kent State University, Kent OH 44242, USA}
\address{$^2$ ISIS Facility, STFC Rutherford Appleton Laboratory, Harwell Oxford, OX11 0QX, UK}
\address{$^3$ Oxford University Department of Physics, Clarendon Laboratory, Parks Road, Oxford, OX1 3PU, UK  }
\ead{aschroe2@kent.edu}

\begin{abstract}
Zero (ZF) and longitudinal field (LF)  muon spin relaxation data of the {\it d}-metal alloy Ni$_{1-x}$V$_{x}$ are presented at several vanadium concentrations $x$ below and above the critical $x_c \approx11$~\% where 
long-range ferromagnetic (FM) order is suppressed. 
The clear single precession frequency observed for Ni, as expected for a homogeneous FM, changes to rather damped osciallation with small V substitution at 
$x=4$~\%,  confirming magnetic inhomogeneities caused by the less magnetic V environments in the magnetic Ni matrix. 
 Furthermore, local fields and spatial field distributions can be estimated to characterize different inhomogeneous regimes developing with $x$ in the FM phase of Ni$_{1-x}$V$_{x}$. 
 In  the regime of $x=7-10$~\% a Kubo Toyabe function well describes the low temperature ZF and LF asymmetry data supporting a static Gaussian field distribution. 
Closer to the quantum critical concentration 
a single scale static Kubo Toyabe function with one field distribution is not sufficient to describe the muon relaxation. These data indicate that further changes in spatial distributions and dynamics are evolving as expected  within the critical regime of a disordered quantum critical point.

\end{abstract}

\section{Introduction}

The study of quantum phase transitions (QPT) is a promising route to understand the origin of unconventional properties and in finding new phases in correlated many body systems. Introduction of ``disorder", through e.g. impurities or sample imperfection can significantly change the nature of the QPT. In itinerant system with Heisenberg symmetry a new quantum critical point is predicted with exotic properties such as observable quantum Griffiths effects \cite{vojta05}.  Examples for disordered systems are not rare, but the degree and significance of the  ``disorder"  is often not clear. The binary alloy Ni$_{1-x}$V$_{x}$ shows the obvious signs of such a disordered QPT from bulk investigations \cite{Ubaid10,as10} and is therefore an ideal example for further microscopic studies. 
muSR is ideally suited as spectroscopic and local probe \cite{yaouanc} to reveal the first insights into these disordered magnetic phases as it has been successfully employed for spin glasses \cite{uemuraSG}, reduced moment correlated electron systems like heavy fermions \cite{amato,maclaughlin,adroja} and phase separated systems close to first order transitions like transition metal compounds \cite{uemuraMnSi}.

The  ferromagnetic (FM) order of Ni is rapidly suppressed replacing Ni by V. 
The  critical temperature $T_c$ decreases approximately linearly and vanishes towards a V-concentration of $x_c \approx 11.4\%$ \cite{Ubaid10} as shown in Fig. 1(a). Magnetization measurements show the typical weak itinerant FM response in the FM phase for $x \leq 11\%$, the ``non FM ordered", paramagnetic (PM) phase beyond $x_c$ is not compatible with independent magnetic moments.  Instead, 
the magnetization displays power laws in temperature (T) and magnetic field with non universal exponents which vary with $x$ in a wide regime $x \ge x_c$ \cite{Ubaid10,as10}. This marks the quantum Griffiths phase (GP) (see e.g. in \cite{vojta06}), characterized by a distribution of magnetic clusters with different fluctuation rates.
This phase is distinctly different than in clean QPT, where universal critical exponents are ideally observed only at $x_c$. We suspect that Ni$_{1-x}$V$_{x}$ is a disordered magnet 
but direct evidence is lacking.
Friedel \cite{Friedel58} already proposed that the drastic average magnetic moment reduction (see e.g. $m_s$ in Fig 2(a)) in Ni  introducing V is not homogeneous, as V (with 5 d-electron less than Ni) might reduce severely the spin of all the Ni-neighbors creating large voids in the magnetization density. Early neutron measurements detected a large magnetic moment disturbance in Ni-V alloys \cite{collins65}. Our bulk magnetization measurements  only show (a mean field behavior of) the order parameter of a FM and no clear signs of the disorder.
Only close to $x_c$ towards very low temperatures deviations from a long range FM become apparent,  cluster glass freezing is observed \cite{Ubaid10} which does not show on this linear T-scale and is not discussed here.
Here we use $\mu$SR to display the ``disorder" in Ni$_{1-x}$V$_{x}$, probing  the local fields in the FM phase. We present how to characterize the inhomogeneities in the FM  and how they evolve upon dilution approaching a disordered QCP.

\section{Experimentals}
Polycrystalline samples of the alloy Ni$_{1-x}$V$_{x}$  were prepared as spherical pellets in the concentration range $x=0$ to $x=12.3$~\%  as before \cite{Ubaid10}. The fcc structure remains unchanged with $x$, the lattice constant increases by $1\%$ for $x=15 \%$ \cite{Ubaid10}. The muon data were collected using the Dolly instrument at PSI, using several pellets of each composition wrapped in silver foil, at low temperatures, in zero (ZF) and longitudinal magnetic fields (LF). 

\section{Evolution of disorder in the ferromagnetic state in Ni$_{1-x}$V$_{x}$}
Fig. 2 shows an overview of the muon asymmetry for several Ni$_{1-x}$V$_{x}$ samples with different V concentration $x$ at low temperatures $T$ in zero field. In pure Ni a nearly undisturbed muon precession  is observed as expected for a clean FM with a single frequency of $f=20.2$~MHz or $\omega=127$ rad/$\mu$s 
corresponding to a local field of $B_0=\omega/\gamma= $0.15 T as seen before \cite{Foy} (using the known gyromagnetic ratio of the muon as $\gamma=2 \pi \times$135.5 MHz/T). Upon doping with V these oscillations appear more ``damped", shift to larger time scales and finally disappear beyond the critical concentration in the PM phase. Already for $x \ge 4$~\% only the first minimum remains visible due to a distribution of fields revealing an inhomogeneous FM. 
Some subtle qualitative changes can already be seen in Fig. 2: Note that $x=4$~\% still shows a second minor maximum or ``nose", and that the minimum for $x=11\%$ is rather shallow. 

In order to model the total asymmetry $A(t)$ for all temperatures we use different polarization functions $P(t)$ for the FM component, a simple exponential function for the PM component (approximating effectively electronic spins of clusters and the diluted V nuclear moments) and a small constant background $A_{BG}$ due to the Ag. 
\begin{equation}
A(t)=A_{FM} P(t)+A_{PM} \exp(-\lambda t) + A_{BG}
\end{equation}
\begin{equation}
P_{osci}(t; \omega,\Gamma,\lambda_L) =\frac{1}{3} \exp(-\lambda_L t)+\frac{2}{3} \cos(\omega t+\phi) \exp(-\Gamma t) 
\end{equation}
\begin{figure}[h]
\begin{minipage}{19pc}
\includegraphics[width=16pc]{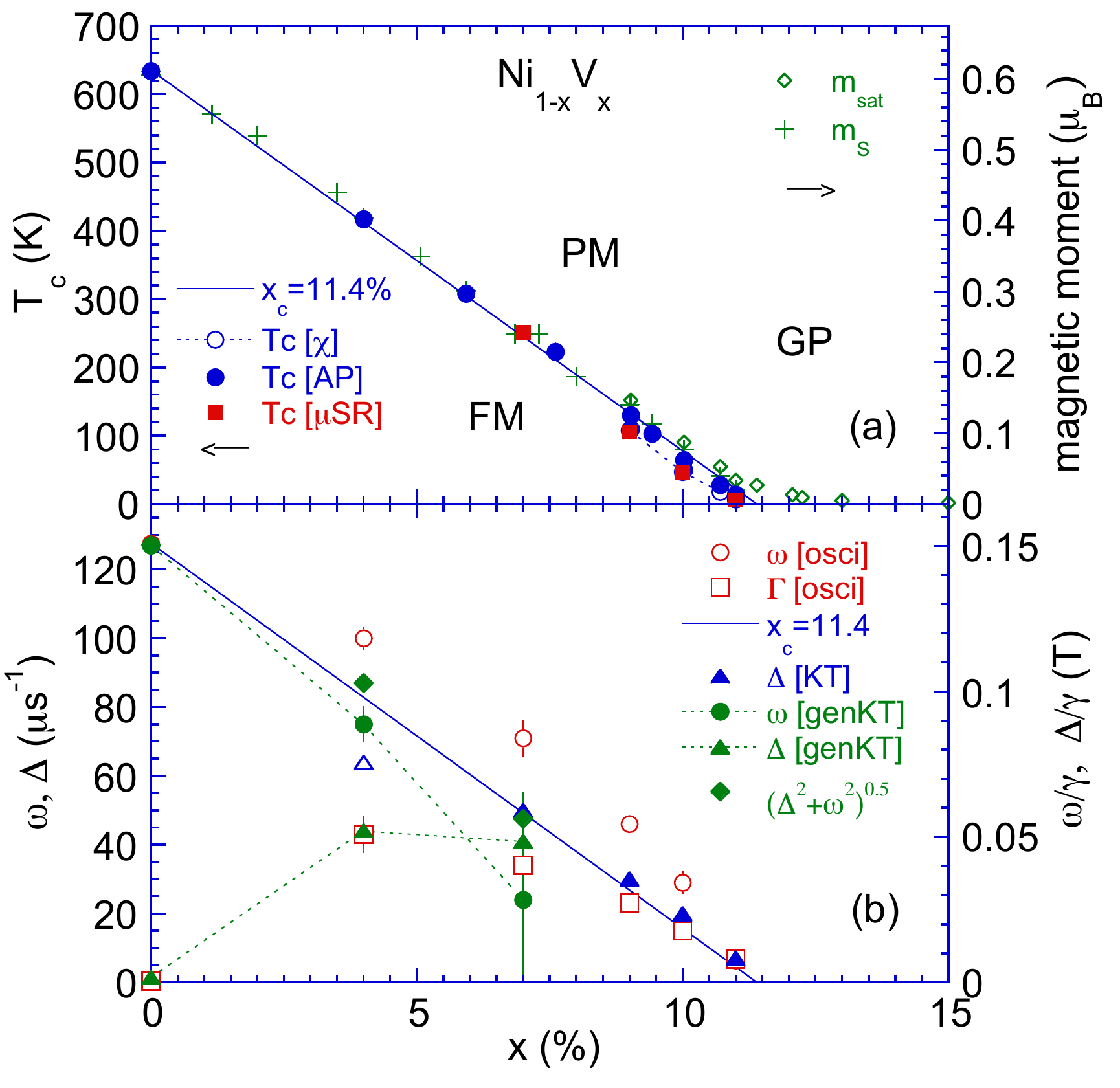}
\caption{\label{label} (a) Phase diagram for Ni$_{1-x}$V$_{x}$: critical temperature $T_c$ vs. V-concentration $x$, separating ferromagnetic (FM), paramagnetic (PM) and quantum Griffiths phases (GP). $T_c$ from magnetization (AP, $\chi$)  \cite{Ubaid10} and $\mu$SR.  Magnetic moment  determined at high ($m_{sat}$) and low fields ($m_s$). Straight line indicates $x_c=11.4$~\%. (b) Angular frequency or local field scales characterizing the center ($\omega$) and distribution width ($\Delta$)  from various $P(t)$ models vs. $x$.}
\end{minipage}\hspace{1.5pc}%
\begin{minipage}{17pc}
\includegraphics[width=16pc]{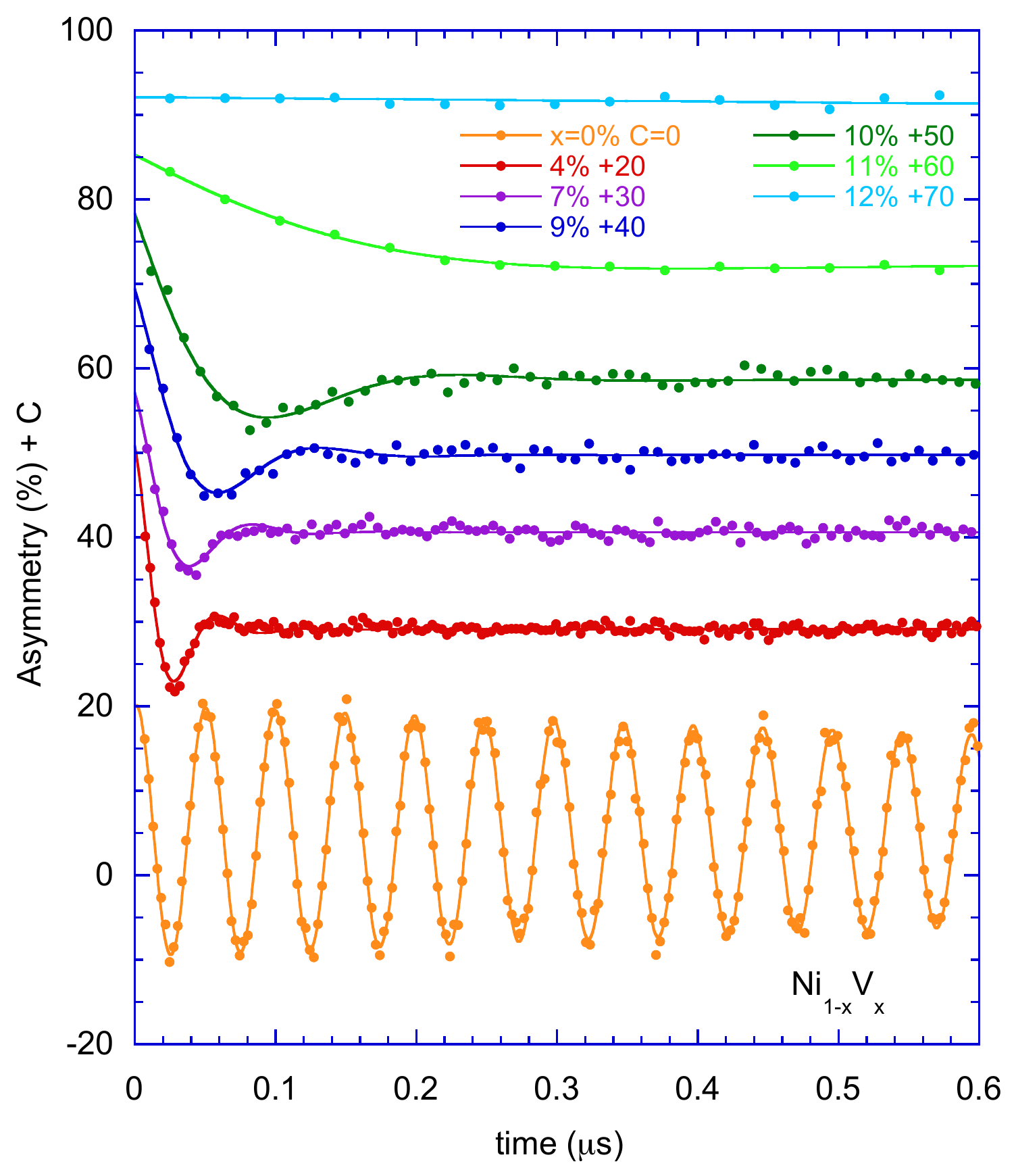}
\caption{\label{label} Asymmetry of Ni$_{1-x}$V$_{x}$ vs time for different V-concentrations $x$, ranging from $x=0$ to $x=12.3$~\% at low temperatures (1.5K $ < T < T_c/4$). Each is shifted by a constant $C$ for clarity. The line represents a fit to the oscillator model with $P_{osci}$.}
\end{minipage} 
\end{figure}

First, we apply an oscillator model $P_{osci}$ (see lines in Fig. 2), a simplified ``damped" expansion of a single precession frequency as expected in a perfect clean FM polycrystal like Ni with one preferred stopping site.  This form is often used and also provides here a parametrization for all $x\le11$~\%. For low $T$, $A_{PM}$ and $\lambda$ are essentially zero, later we need to set $\lambda_L=\lambda$ for all $T$. Also we keep $\phi=0$ for higher $x$ to produce consistent parameters. The main frequency $\omega$  decreases with $x$ as shown in Fig. 1(b). The damping constant $\Gamma$ or transverse field depolarization rate is increased for $x=4$~\% then decreases further with $x$. 
$\Gamma$ can be seen as a measure of field distribution, as $\omega$ is indicating some local field, but comparison to other models like $P_{genKT}$ (see eq.(4) below) show that in the ``damped" cases $\omega$ is not exactly the mean field nor the field distribution.   $\Gamma$ has half the value of $\omega$ at lowest T for all samples with $7\%<x<10$~\%, which marks the regime with a Gaussian field distribution as discussed below.

More revealing simple models can be applied, but are restricted to certain $x$-regimes. In effect, for $x=7-10$~\% a static Kubo-Toyabe (KT) function $P_{KT}$ describes the data well for low T, indicating a Gaussian distribution of local fields \cite{Hayano} in the FM state. 
\begin{equation}
P_{KT}(t; \Delta,\nu =0) =\frac{1}{3} +\frac{2}{3} (1-(\Delta t)^2) \exp(-\frac{1}{2}(\Delta t)^2) \\
\end{equation}
The single parameter, the width $\Delta$ 
of the distribution decreases with $x$ towards  $x_c$ as shown in Fig. 1(b). This field distribution follows the magnetic moment $m_s$ on the same straight line in Fig.1(a,b). 
It is remarkable that inside a FM a ``random" magnetic field distribution is found as originally shown for randomly oriented frozen moments \cite{Hayano}. We suspect here that the distribution of different local fields indicates a spatial variation of magnetic moments in a long range ordered FM state. 
These muon data provide evidence that this system is indeed inhomogeneous or ``disordered" as required to show this disordered QCP with distinct properties. 

A Kubo Toyabe fit $P_{KT}$ is possible for $x=4$~\% but can be improved by a generalized form $P_{genKT}$ \cite{Kornilov} allowing a finite center field $H_0=\omega/\gamma \neq0$ from large domains or crystallites which is the center of a Gaussian distribution with width $\Delta H=\Delta/\gamma$. 
\begin{equation}
P_{genKT} (t; \Delta,\omega,\nu=0)= \frac{1}{3} +\frac{2}{3} \exp(-\frac{1}{2}(\Delta t)^2) [\cos(\omega t) - \frac{1}{\omega}(\Delta t)^2  \sin(\omega t)]
\end{equation}
The two parameters $\omega$ and $\Delta$ are shown in Fig. 1(b).  $\Delta$ is similar for both $x=4$~\% and $x=7$~\%, overall the maximum field distribution appears at about  $\frac{1}{2}x_c$. The characteristic field $H_0$ in $x=4$~\% is still high due to intact magnetic domains,  while $H_0$ is rather close to zero in $x=7$~\%, the field distribution $\Delta H$ dominates. The transition from dominating FM domains to random distribution occurs in between these concentrations at about $\frac{1}{2}x_c$. That matches the concentration $x=5.6$~\% where the number of reduced moment Ni-atoms (due to one or more V neighbors) exceeds the number of undisturbed ``magnetic" Ni-atoms (with only Ni neighbors) in a fcc lattice. The local muon probe distinguishes the different degrees of inhomogeneities in this FM introduced by V substitution. The model $P_{genKT}$ provides even a direct quantitative assessment of a internal field distribution width $\Delta/\gamma $ and a center local field $\omega/\gamma$. An effective field $\sqrt{\Delta^2+\omega^2}/\gamma$ finally scales with the mean magnetic moment $m_s(x)$.

The KT function $P_{KT}$ does not work as well towards the more diluted regime approaching $x_c$, the minimum becomes very shallow, for e.g. $11$~\%, questioning the long range order of this compound.
Beyond $x_c$, for $x=12.3$~\%, a small depolarization rate ($\lambda=0.07\mu s^{-1}$ for 2K) describes the data in ZF. Extended time, more temperature and magnetic field data and analysis  are required to reveal more details about the fluctuation rates beyond the FM regime.

\section{Contrasting the field distributions in Ni$_{1-x}$V$_{x}$ with $x=9$~\% and $x=11$~\%}
The asymmetry spectra for $x=9$~\% can be modeled well by a KT function. For higher temperature we used a dynamic KT approximation with finite fluctuation rates $\nu$  \cite{Hayano,WIMDA} for the FM contribution (as presented in Fig. 3(a)). 
Application of a longitudinal field (LF) at low T  confirms the main parameter choices of a KT description (see Fig.3(b)).
The straight lines shown are a KT fit with unchanged parameters from ZF with internal field = external LF, better fit results are achieved with reduced internal fields in these FM spherical samples. 
Fig.4 (a-d) presents the T-dependence of the fit-results of the exploratory oscillator model and the KT model. Panel (a) shows the main field distribution $\Delta$ together with $\omega$ and $\Gamma$. 
$\Delta (T)$ follows nearly
a mean field T-dependence like the magnetic moment $m_s(T)$ vanishing at $T_c$. The fixed ratio $\omega/\Gamma$ of 2 (observed for several $x$ at low T) changes towards higher T  towards 1, before $\omega$ vanishes. The ferromagnetic amplitude ratio, the ferromagnetic contribution divided through the total  contribution, $A_{FM}/(A_{FM}+A_{PM})$, increases rapidly below $T_c$ and then saturates to about 1 independent of the model (see Fig.4(b)). The depolarization rate $\lambda$ is quite small in the PM regime and also very small, close to the detection limit, in the FM for $T<0.8 T_c$ (as $\lambda_L$) in $P_{osci}$. This rate peaks towards $T_c$, indicating some critical dynamics. The dynamic KT model reveals directly critical fluctuations increasing rapidly at $T_c$ (see Fig.4(d)). We see that below about $0.8T_c$ magnetic fluctuations become irrelevant.
All quantities as $A_{AF}$, $\Delta$, $\nu$ indicate a critical temperature of $T_c=108$~K consistent with low field susceptibility $\chi$ measurements \cite{as10}, a bit lower than the usual high field Arrott plot (AP) determination of $T_c=130$~K.
\begin{figure}[h]
\begin{minipage}{20pc}
\includegraphics[width=18pc]{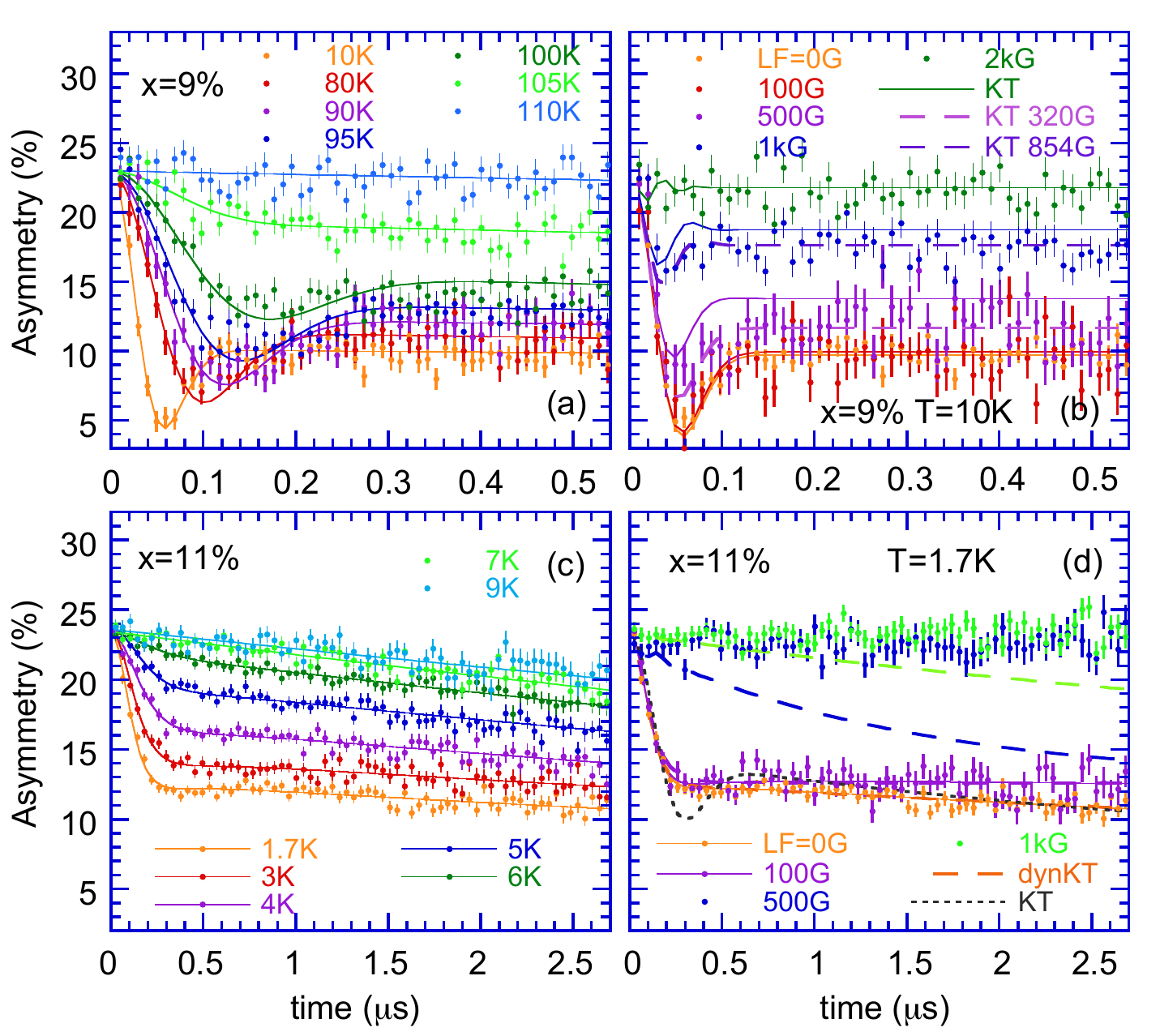}\hspace{2pc}%
\caption{\label{label}(a) ZF spectra for Ni$_{1-x}$V$_{x}$ with $x=9~\%$. Lines from $P_{KT}$. (b) LF spectra for $x=9$~\% at $T=10$~K. Lines from $P_{KT}$: field = external LF (solid), adjusted internal field (dashed). (c) ZF spectra for $x=11$~\%. Lines from $P_{GBG}$. (d) LF spectra for $x=11$~\%. Dotted line: static $P_{KT}$. Dashed lines: dynamic $P_{KT}$. Solid lines: $P_{GBG}$.}
\end{minipage}\hspace{1.5pc}
\begin{minipage}{16pc}
\includegraphics[width=16pc]{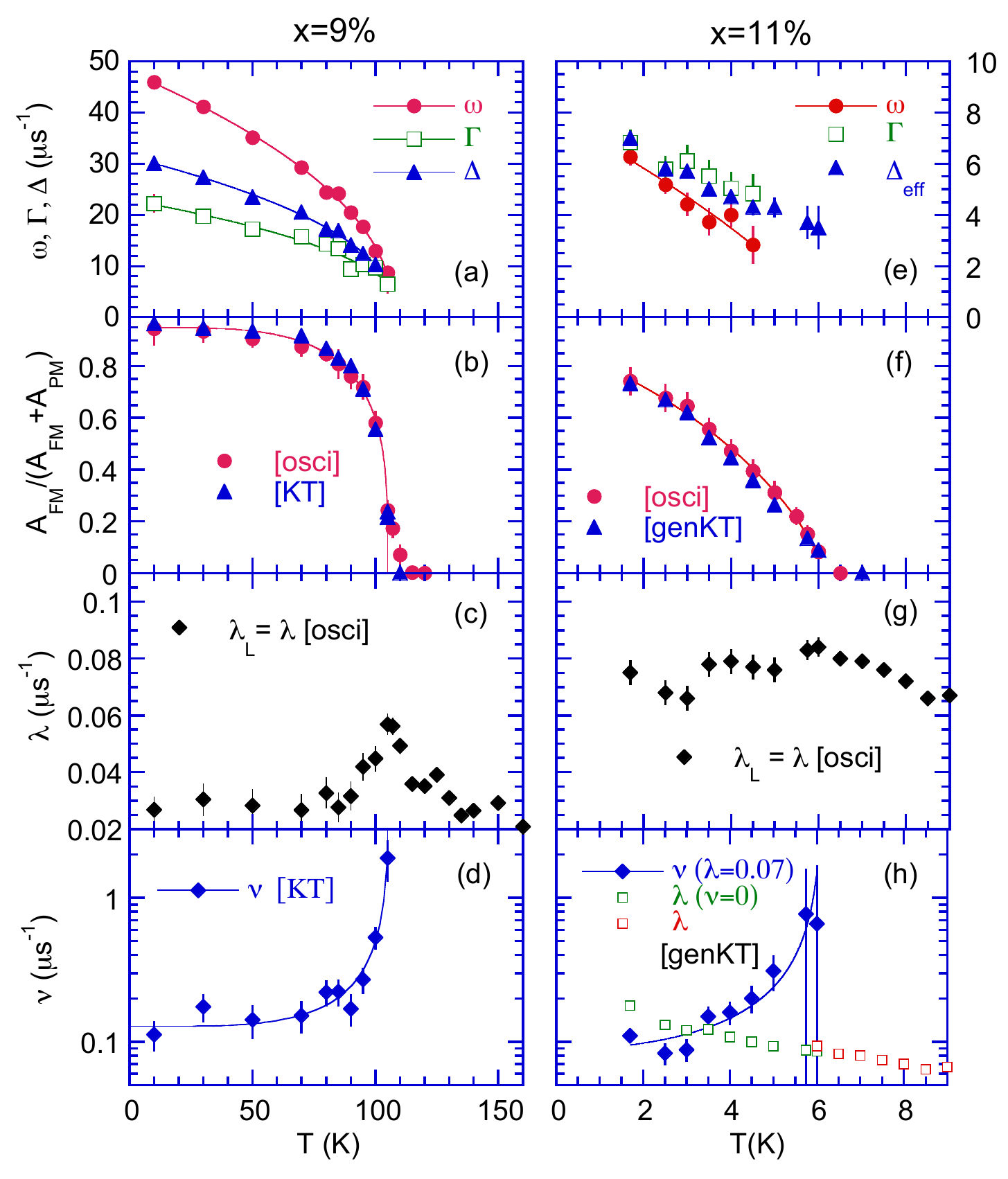}\hspace{2pc}%
\caption{\label{label} T-dependence of various fit parameters using $P_{osci}$, dynamic $P_{KT}$ or $P_{GBG}$ to model $A(t)$ as shown in Fig. 3 for $x=9$~\% (a-d) and for $x=11$~\% (e-h). Lines are guide to the eye.}
\end{minipage}
\end{figure}

Magnetization measurements indicate that $x=11$~\% is still a ferromagnet displaying Arrott plots and no indication of a cluster glass transition is found in magnetic susceptibility.  A strong ZF  depolarization is observed, but the minimum is too shallow for the static KT form even at lowest T (see Fig. 3(d)).  
The ZF data can be modeled quite well applying a {\it dynamic} KT function ($P(t,\Delta,\nu>\Delta)$), using the strongly fluctuating approximation of a Keren function \cite{Keren,WIMDA}. But the present LF data (e.g. in 500G) cannot be reproduced keeping the same fluctuation rate as in ZF (see dashed line in Fig. 3(d)). More intermediate fields are needed to study field dependent changes.  We alternatively applied a {\it static} distribution function which describes multiple field distributions, narrow and wide ones 
to increase the weight towards very small field regions. Taking the response  of many Gaussian distribution of fields with different $\Delta's$ following a Gaussian distribution with width $\Delta_W$ and center $\Delta_0$ gives the so called simple Gaussian broadened Gaussian function $P_{GBG}$ \cite{noakes,noakesmod}. 
\begin{equation}
P_{GBG}(t;\Delta_{eff},R,\nu)= \frac{1}{3} \exp(-\frac{2}{3}\nu t)+\frac{2}{3} ((1+R^2)/N)^{3/2}(1-(\Delta t)^2/N) \exp(-\frac{1}{2N}(\Delta t)^2)
\end{equation}
with $N=1+R^2+R^2\Delta^2t^2$, where $R=\Delta_w/\Delta_0$ and $\Delta^2=\Delta_{eff}^2=\Delta_0^2+\Delta_w^2$.
This form describes the data well as shown in Fig. 3(a) (with R=0.7). Note that an additional PM contribution with a small depolarization rate is always present even for low T. The effective field distribution width $\Delta_{eff}$ is shown in Fig. 4(e) together with the $P_{osci}$ parameters.  They decrease towards higher $T$, a sharp critical temperature  $T_c$ cannot be determined. Also, the ratio of these parameters at $x=11$~\% is different than those at $x=9$~\% at low T, more similar to those at $x=9$~\% close to $T_c$. The FM amplitude ratio increases only gradually towards low T (Fig. 4(f)), not as sharp as in $x=9$~\%. At the lowest T,  only $75$~\% of the amplitude can be described by the FM component $P(t)$, while an extra (rest of $25$~\%) exponential ``paramagnetic" contribution remains, independent of the model. The depolarization rate $\lambda$ (=$\lambda_L$) shows barely a maximum at $T_c$ and  does not decrease towards low values as for $x=9$~\% in the oscillator model (Fig.4(g)). Keeping $P_{GBG}$ static ($\nu=0$) the depolarization rate $\lambda$ of the PM contribution increases further below $T_c$ (Fig.4(h)). A small LF=100G suppresses these fluctuations (see Fig.3(d)), while the steep FM depolarization seems unaffected. Alternatively, allowing small fluctuations ($\nu >0$) and keeping $\lambda=0.07 \mu s^{-1}$ constant produces some critical dynamic towards $T_c$.  Estimates of $T_c \approx 6$~K from these muon data is in agreement with low field susceptibility $\chi$ measurements \cite{as10}, much lower than Arrott plot (AP) extrapolations of $T_c =14$~K. \cite{Ubaid10}

While the inhomogeneous magnetism in $x=9$~\% could be well characterized by one scale, the KT distribution width $\Delta$, multiple field distributions seemed better in $x=11$~\%. Furthermore, a PM contribution due to fluctuating clusters never vanishes at low T. These models are far too simple but indicate that fluctuating moments and the spatial inhomogeneities further evolve towards $x_c$.  
$11$~\% is not just a scaled down version of $9$~\% with a smaller field range. 

\section{Conclusion}
Ni$_{1-x}$V$_{x}$ offers the opportunity to study a FM with different degrees of ``spatial disorder" or inhomogeneities, introduced through V substitution from a pristine FM to a disordered QCP. Analyzing the muon asymmetry with a generalized KT model, $P_{genKT}$, the mean local field and the local field distribution could be determined for a large $x$ regime. These parameters measure the degree of ``disorder" and allow a distinction between different regimes within the FM state. For lower $x$ intact domains with finite fields dominate, while for larger $x$ the field distribution dominates. This distribution width is the dominating energy scale in a wide region 7~\% $ \leq x\leq10$~\%, it follows the order parameter, $m_s$. Approaching the critical concentration, $x_c$, simple single scale fit functions cease to work. Multiple distributions with multiple time scales evolve at the edge of the Griffiths phase, dominated by clusters with different time scales. 
So far we only used simple models to probe local field distributions in the FM phase. To relate this information to the spatial variation of inhomogeneities and correlation length a microscopic model is needed. 
The simple fcc lattice of Ni-V looks promising as a good model system.
The study of the complex scenario in the quantum critical regime and towards the quantum Griffiths phase is a further challenge in this {\it d}-metal alloy.

\section{Acknowledgment}
We thank S. Ubaid-Kassis for the original samples and W. Hayes for helpful discussions. This work was supported in part by OBR Research Challenge from KSU and part of this work was carried out at the Swiss Muon Source, Paul Scherrer Institute, CH.

\section*{References}

\end{document}